# Towards the Probabilistic Earth-System Model[1]

by

T.N. Palmer▲*, F.J. Doblas-Reyes*, A. Weisheimer*, G.J. Shutts*‡,

J. Berner§, J.M. Murphy‡

Affiliations:

*ECMWF, Shinfield Park, Reading, Berkshire, RG2 9AX, UK

§NCAR, P.O. Box 3000, Boulder, CO 80307-3000, USA

‡ Met Office, FitzRoy Road, Exeter, Devon, EX1 3PB, UK

▲ corresponding author: Tim.Palmer@ecmwf.int

3 December 2008

---

[1] Based on a presentation at the WCRP/WWRP/IGBP World Modelling Summit for Climate Prediction

# Abstract

Multi-model ensembles provide a pragmatic approach to the representation of model uncertainty in climate prediction. On the other hand, such representations are inherently *ad hoc*, and, as shown, probability distributions of seasonal climate variables, made using current-generation multi-model ensembles, are not accurate. Results from seasonal re-forecast studies suggest that climate model ensembles based on stochastic-dynamic parametrisation are beginning to outperform multi-model ensembles, and have the potential to become significantly more skilful than multi-model ensembles.

The case is made for stochastic representations of model uncertainty in future-generation climate prediction models. Firstly, a guiding characteristic of the scientific method is an ability to characterise and predict uncertainty; individual climate models are not currently able to do this. Secondly, through the effects of noise-induced rectification, stochastic-dynamic parametrisation may provide a (poor man's) surrogate to high resolution. Thirdly, stochastic-dynamic parametrisations may be able to take advantage of the inherent stochasticity of electron flow through certain types of low-energy computer chips, currently under development.

These arguments have particular resonance for next-generation Earth-System models, which on the one hand purport to be comprehensive numerical representations of climate, but where on the other hand, integrations at high resolution may be unaffordable.

# 1. Introduction

Physical climate models evolved out of numerical weather prediction models, from a necessity to include representations of long-timescale physical processes. In turn, Earth-System models (ESMs) are now evolving out of physical climate models from a need to include representations of important biogeochemical and exospheric processes. The ESM is often defined as an attempt at a comprehensive numerical algorithm for simulating and predicting the evolution of Earth's climate.

A guiding principle of the scientific method is an ability to characterise and predict uncertainty. Estimates of uncertainty in predictions of climate change are critical in guiding both mitigation policy and adaptation strategies on climate. Hence, if an ESM purports to be both scientific and comprehensive, it should be capable of predicting uncertainties in its own predictions.

In practice, however, this is not the case. Instead, it is conventional to estimate forecast uncertainty by pooling together output from different climate models in the form of a multi-model ensemble, hereafter MME (e.g., Palmer and Räisänen, 2002; Giorgi and Mearns, 2003; Tebaldi et al., 2004; Weisheimer and Palmer, 2005; Greene et al., 2006).

The collaborative spirit engendered by the MME concept could be seen as a virtue. On the other hand, MMEs are, by their nature, *ad hoc*; there is no premeditated effort by the modelling community to ensure that a MME properly samples the relevant uncertain directions in state space. Indeed, as argued below, it is unlikely one could design a MME to do this, even in principle. As a result, MMEs are sometimes referred to pejoratively as "ensembles of opportunity".

Despite this, MMEs do provide more skilful seasonal climate forecasts than single model predictions (Palmer et al., 2004). However, this does not imply that probability distributions of climate variables derived from contemporary MMEs are themselves accurate. As shown in Section 2, such distributions can in fact be quite inaccurate and in practice this has necessitated the use of empirical bias correction in order to perform skill assessments. However, since climate is a profoundly nonlinear system, such linear bias corrections cannot guarantee reliable probability forecasts.

As an alternative to the MME, stochastic-dynamic parametrisation (Palmer, 2001) has been developed in numerical weather prediction to represent model uncertainty in single model ensembles. In Section 3, ensembles of single models with stochastic-dynamic parametrisation are compared with the MME in seasonal climate prediction. It is found that the performance of such schemes, particularly when combined together, is beginning to be competitive with the MME.

In Section 4, we therefore put forward a three-part thesis that next-generation climate models should be explicitly probabilistic. Firstly, as discussed above, a guiding characteristic of the scientific method is an ability to characterise and predict uncertainty and individual climate models are not currently able to do this. Secondly, through the effects of noise-induced rectification, stochastic-dynamic parametrisation may in some respects act as a (poor man's) surrogate of high resolution. Thirdly, stochastic-dynamic parametrisations may be able to take advantage of the inherent stochasticity of electron flow through certain types of low-energy computer chips.

These arguments have particular resonance for next-generation Earth-System models, which on the one hand purport to be comprehensive numerical representations of climate, and where on the other hand, integrations at high resolution may be unaffordable.

Conclusions are given in Section 5.

## 2. The Accuracy of Climate Probability Distributions Derived from Multi-Model Ensembles.

MMEs are used in seasonal climate forecasting and have been shown to outperform ensembles of single deterministic models in terms of probabilistic skill scores. This implies that it is necessary to include estimates of model uncertainty in some form or another, in climate prediction ensembles. However, this result does not imply that MMEs do sample model uncertainties adequately. The analysis in Palmer et al (2004) for example, was performed after a linear bias correction had been applied. As discussed below, linear bias correction is not sufficient to guarantee reliable forecasts on climate change timescales.

To illustrate the essential unreliability of the MME, Figure 1 shows two schematic probability density functions (PDFs) of seasonal-mean climatic variables (e.g. surface temperature). The PDF in Figure 1a is presumed to have been determined from observations, whilst Figure 1b is a comparable PDF derived from a hypothetical MME. The observed distribution is divided by two tercile thresholds (black dashed lines). Hence, the blue area in Figure 1a is precisely one third of the total area under the curve. These terciles from the observed distribution are also used to divide the MME PDF. If the MME PDF is accurate, the blue area in Figure 1b would also equal one third. The extent to which the blue area is not equal to one third is a measure of unreliability used in this paper.

## 2.1 The IPCC AR4 multi-model ensemble

An MME of simulations of 20$^{th}$ Century climate was carried out for the IPCC AR4 using the latest generation of coupled atmosphere-ocean climate and Earth System models (Weisheimer and Palmer, 2005). Here we analyse 18 member MME PDFs of seasonal near-surface temperature during the 20-year period from 1971 to 1990, and compare them with PDFs derived from ERA-40 (Uppala et al., 2005) to estimate the MME adequacy, or reliability, in the above discussed sense.

Figure 2a shows a map of the IPCC MME frequency of lying below the observed lower tercile for June - August (JJA) near-surface temperature for all gridpoints over the globe (interpolated to a common T42 grid). If an ideal MME were to be reliable, 1/3 of the data sample would fall below that tercile. In our case (with 18x20 data points), the MME frequency $f_{MME}$ is not statistically different from this reference threshold of 1/3 if $0.247 \leq f_{MME} \leq 0.425$, with 99% confidence. In Fig 2, areas where $f_{MME}$ is not statistically different from 1/3 are shown white. Ideally, 100% of the area of the globe should be shown white. In fact, only 16% of the globe is so indicated. For the other 84%, there is poor agreement between the observed and modelled frequencies. For example, it can be seen that for many parts of the Northern Hemisphere the agreement is exceptionally poor ($f_{MME} > 0.7$; coloured deep blue). By contrast, for the cold upwelling regions on the western coasts of South America and Southern Africa, $f_{MME} < 0.1$.

Figure 2b shows maps giving the MME frequency of lying below the lower tercile for December - February (DJF) temperature. Similar to JJA conditions, $f_{MME}$ is much too large in tropical and subtropical regions, and too small for parts of the extratropics. $f_{MME}$ is accurate for only 19% of the globe.

## 2.2 How reliable is linear bias correction in non-linear systems?

The diagnostics in Figure 2 show that, for DJF and JJA seasonal means, the area below the lower tercile is significantly biased across most of the globe. This is an indication more generally that the MME PDFs are not accurate, and hence that the pool of models used in the MME is not properly sampling true model uncertainty.

This indicates that it is certainly desirable to empirically bias-correct model output, e.g. before disseminating results to users, and weather and seasonal forecast studies clearly show benefit from such empirical bias correction. On the other hand, whilst bias correction may be necessary to improve the accuracy of forecast PDFs, it is not sufficient. An example of the limited adequacy of bias correction in a simple nonlinear model illustrates this basic point.

Consider the Lorenz (1963) equations

$$\begin{aligned} \dot{X} &= -\sigma X + \sigma Y \\ \dot{Y} &= -XZ + rX - Y \\ \dot{Z} &= XY - bZ \end{aligned} \qquad (2.1)$$

Let us refer to (2.1) as the SYSTEM, and label the two regimes of the SYSTEM as "westerly" and "blocked". Let the state-space coordinates of these regime centroids be $(X_0, Y_0, Z_0)$, $(-X_0, -Y_0, Z_0)$. Figure 3a shows the attractor of the Lorenz (1963)

SYSTEM in the X-Y plane. Now consider a MODEL of the SYSTEM, whose attractor is defined by rotating the Lorenz attractor by $\pi/4$ radians about a line parallel to the Z axis, through the centroid of the SYSTEM's "westerly" regime (Figure 3b). As a consequence of this, the MODEL attractor's "westerly" regime centroid agrees well with the SYSTEM attractor's "westerly" centroid, but the MODEL attractor's "blocked" regime centroid is poorly represented. As a consequence, the climatic mean state of the MODEL is manifestly biased with respect to that of the SYSTEM.

Imagine now that the SYSTEM is subject to a "climate change" given by the prescribed forcing $\mathbf{f}_0 = (f_0, f_0, 0)$ so that, under this "climate change" we have

$$\begin{aligned}\dot{X} &= -\sigma X + \sigma Y + f_0 \\ \dot{Y} &= -XZ + rX - Y + f_0 \\ \dot{Z} &= XY - bZ\end{aligned} \quad (2.2)$$

Here we choose $f_0 = 4$. Both the responses of the SYSTEM and the MODEL to the imposed forcing are given in Figure 3c. The MODEL response has been illustrated in "bias-corrected" format, so that the tails of the arrows, representing the time-mean state of the unforced SYSTEM and unforced MODEL, are both located at the same point in X-Y state space.

It can be seen that, bias correction notwithstanding, both the magnitude and (state-space) direction of the imposed forcing is quite different in the MODEL and the SYSTEM. (In the case shown, the MODEL's response to the imposed forcing is weaker than the SYSTEM's response). This can be readily understood. As shown in Palmer (1999), the response to the imposed forcing $\mathbf{f}_0$ in the Lorenz (1963) SYSTEM, does not necessarily lie along the direction of forcing; a substantial component of the response is associated with a change in the frequency of occurrence of the regimes, irrespective of the direction of forcing. When $\mathbf{f}_0$ is not parallel to the line joining the regimes, the response lies in a direction somewhere between $\mathbf{f}_0$ and the line joining the regimes. The largest response occurs when $\mathbf{f}_0$ is parallel to the line joining the regimes.

Figure 3c illustrates the fact that linear bias correction cannot be assumed to correct for inherent model deficiencies when estimating the response of a nonlinear system to some imposed forcing, e.g., to climate change. The example used is clearly only illustrative, but there is evidence for the existence of regional regimes in the real climate system (e.g., Straus et al, 2007).

In summary, probability distributions from contemporary climate model MMEs are not accurate and therefore do not represent model error uncertainty adequately. Empirical bias corrections to these probability distributions are necessary to improve accuracy, but such linear bias-correction techniques cannot be assumed sufficient to ensure accuracy in a nonlinear system such as climate. Since climate change predictions are now being used to guide regional multi-billion dollar infrastructure investments for society to adapt to climate change, it is essential that PDFs of climate change are as accurate as possible. In the next section, we therefore consider possible alternatives to the MME; one such alternative is now beginning to challenge the MME.

# 3. Comparing Multi-Model and Stochastic-Dynamical Model Ensembles in Seasonal Forecast Mode.

Stochastic-dynamic parametrisations have been developed as an alternative to the MME (Palmer, 2001). The original motivation for developing these stochastic parametrisation schemes was to represent model uncertainty in numerical weather prediction (Buizza et al., 1999). More recently, these schemes have been incorporated into seasonal forecast systems (Berner et al., 2008). A systematic analysis of these schemes on the climate-change timescale has yet to be performed.

In this section, we compare MMEs against stochastic-dynamic ensembles, based on a coordinated set of seasonal-timescale re-forecasts made as part of the European Union ENSEMBLES project. Although a comparison on seasonal timescales is not necessarily a reliable guide to an assessment of their relative performance on longer climate-change timescales, there is a growing belief that the constraints and insights of numerical weather prediction and seasonal forecasting can be brought to bear on the climate problem, through the concept of seamless prediction (Palmer and Webster, 1995; Rodwell and Palmer, 2007; Palmer et al., 2008).

## 3.1 Experimental set-up

Sets of ensemble re-forecasts over the period 1991 to 2001 have been performed (Doblas-Reyes et al., 2009). The simulations are started on 1$^{st}$ of May and 1$^{st}$ November (00 GMT) and run for seven months.

The coupled forecast systems that contributed to this experiment are as follows:

- The MME comprises: IFS/HOPE (ECMWF), ARPEGE/OPA (Météo-France), GloSea, DePreSys_ICE (both UK Met Office) and ECHAM5/MPIOM (IfM-GEOMAR Kiel). The MME ensembles have 45 members (9 from each model member). .

- The Stochastic Tendency (Buizza et al, 1999) and Cellular Automaton Backscatter Scheme (CASBS; Shutts 2005, Berner et al, 2008) scheme have been implemented in the ECMWF forecast system (IFS/HOPE) and run with nine-member ensembles.

- The Perturbed-Parameters technique has been implemented in the UK Met Office forecast system (DePreSys_PPE) and run with nine different versions of HadCM3.(Smith et al, 2007).

More details on the experimental design are available in Doblas-Reyes et al. (2009).

## 3.2 Comparison between forecast systems

The skill of the different ensemble systems will be studied using standard probability scores based on dichotomous events. The Brier skill score (Jolliffe and Stephenson, 2003; BSS henceforth) with respect to climatology has been used as the measure of forecast quality. The BSS is a measure of the relative benefit of the forecasts with respect to using the naïve climatological probabilities and is defined as BSS=1-BS/$BS_c$, where $BS_c$ is the Brier score of the climatological forecast, the one that always issues as forecast probability the historical frequency of the event. Forecast quality measures are computed taking into account the systematic error of the forecast systems. This means that the threshold that defines the forecast event is chosen separately for the verification dataset and the set of forecasts, considering the re-forecast values for all the available years for the same start date and lead time.

Figure 4a shows scatter plots of the BSS that compares the forecast quality of the stochastically-parametrised ensemble with the forecast quality of an equally-sized (9-member) MME. Scores for five variables over the tropical band, the Southern extratropics and Northern extratropics. Predictions for the two different starts dates (May 1st and November 1$^{st}$), two forecast periods (seasonal averages with one and three month lead time) and three different events (anomalies above the upper tercile, above the median, and below the lower tercile) have been plotted together (a total of 180 cases).

The first point to note is that a version of the MME with nine members (that draws members from each of the five single-model ensembles contributing to the MME) performs more often better than the stochastically parametrised ensemble. This is particularly obvious for the tropics. For example, the MME is significantly better (with 95% confidence) than the ECMWF stochastically-parametrised physics ensemble 24 times, whilst the ECMWF stochastically-parametrised physics system is better than the MME only three times.

The superiority of the MME in terms of BSS is largely part a result of an increased reliability of its probability forecasts (not shown) and this becomes even more substantial when the multi-model is used with its full available ensemble size (45 members).

Probability forecast unreliability is a characteristic of forecast quality that may reflect an underdispersive, ie overconfident ensemble. That is to say, the results may indicate that the current stochastic parametrisations are simply too conservative in representing model uncertainty. To test this hypothesis, the spread of the stochastically parametrised ensembles has been statistically re-calibrated using the calibration method described in Doblas-Reyes et al. (2005). The calibration modifies empirically the ensemble spread, in order to reduce the gap between the root mean square error of the ensemble mean forecast, and the ensemble spread as measured in terms of standard deviation of the ensemble members around the ensemble mean. The method assumes that the standard deviation of the re-calibrated prediction is the same as that of the reference and that the potentially predictable signal after re-calibration is equal to the correlation of the ensemble mean with the observations. This implies that if a re-calibrated ensemble member can be expressed as

$$z_{ij} = \alpha x_i + \beta x_{ij} \tag{3.1}$$

where $x_i$ is the ensemble mean for time step i and $x_{ij}$ is the difference of ensemble member j with respect to $x_i$, the coefficients α and β are given by

$$\alpha = \rho \, s_r / s_m \qquad \beta = s_r \frac{\sqrt{1-\rho^2}}{s_e} \tag{3.2}$$

In these equations, $s_e$ is the standard deviation of all $x_{ij}$ (the mean spread), $s_m$ the standard deviation of the original ensemble mean, $s_r$ the standard deviation of the observations, and ρ the time correlation between the observations and the original ensemble-mean forecast. While in a forecasting context this re-calibration method would be applied in cross-validation to avoid overfitting of the forecast correction, we here apply it in-sample to obtain an indication of the potential for improving skill of an ensemble forecast system based on stochastic parametrisation, but with improved reliability due to less conservative stochastics.

Figure 4b shows the BSS scatter plot for the nine-member multi-model ensemble versus the perfectly calibrated stochastically parametrised ensemble. In this case, the stochastically parametrised ensemble performs more often better than the multi-model. The results are dramatic - out of 180 possible cases, the re-calibrated stochastically parametrised ensemble outperforms the MME with 95% confidence on 112 occasions, whilst the MME only outperforms the stochastically parametrised ensemble twice. The re-calibrated stochastically parametrised ensemble does not show any negative values of the BSS.

In order to interpret this latter result, let us return to some basic discussion of the nature of parametrisation uncertainty. A useful delineation of such uncertainty is in terms of uncertainty in the optimal form of the bulk-formula representation of the parametrisation, referred to here as B-uncertainty, and structural uncertainty in the use of bulk-formula representations, referred to here as S-uncertainty. For example, the DePreSys perturbed parameter scheme addresses B-uncertainty, whilst the CASBS scheme in the IFS/HOPE model addresses S-uncertainty (see discussion in Section 4.1 below). The stochastic tendency method is perhaps more directly associated with B-uncertainty, but, unlike the perturbed parameter approach the stochastic tendency perturbations are incoherent on temporal scales of days and longer, whilst the perturbed parameter perturbations are perfectly coherent (in parameter space) over all timescales.

One hypothesis for the success of the re-calibrated stochastically parametrised ensemble, is that the IFS/HOPE stochastic model is not sufficiently representative of both B- and S- uncertainty. To assess this, the skill of an ensemble which combines the uncalibrated DePresSys and IFS/HOPE stochastically parametrised ensemble has been compared with the MME. (It can be noted that results of a comparison of DePresSys vs MME is similar to that for stochastically parametrised ensemble vs MME, as shown in Figure 4b.)

Hence Figure 4c shows the scatter plot for the BSS of a combination the stochastically parametrised ensemble and perturbed-parameters ensembles (18 members in total) with an 18-member MME. It is interesting to see that, in spite of both the stochastically parametrised ensemble and perturbed-parameters ensembles performing worse than the nine-member MME, when both are put together, the resulting ensemble performs somewhat better than an equally sized MME. For example, out of the possible 180 cases, there are 19 cases where the stochastic ensemble outperforms the MME at the 95% confidence level, and only 9 cases of the converse (results obtained with a two-sample test based on differences of bootstrapped estimates, where the pairs re-forecast/analysis were resampled with replacement 1,000 times). It is true that the stochastic ensemble used in Figure 4c is not based on a single model with stochastic parametrisation, since the perturbed parameter scheme has been implemented in a different model to the stochastic tendency and stochastic backscatter scheme. Nevertheless, the resulting ensemble comprises only two model systems, rather than the 5 models of the MME. Further work is needed to combine the stochastic schemes into one model system.

## 4. Towards the Probabilistic Earth-System Model

Based on results in the previous section, a three-part case is made for next-generation earth-system models to be inherently probabilistic. Both theoretical and pragmatic issues are discussed, the latter relating to matters of computational efficiency.

## 4.1 Representation of Model Uncertainty

Stochastic-dynamic parametrisation provides an explicit representation of model uncertainty , and in this respect is beginning to be competitive with the MME. Above all, stochastic-dynamic parametrisation offers the means to represent model uncertainty in a way which is less *ad hoc* than is the multi-model concept.

As an example of this, consider the type of coarse-grain analysis of cloud resolving models discussed in Shutts and Palmer (2007). By defining the cloud-resolving model as "truth", probability distributions of coarse-grain sub-grid processes can be derived, conditioned on the coarse-grain flow. The extent to which these coarse-grain probability distributions are not Dirac distributions is a measure of the inherently stochasticity of the parametrisation problem, and the structure of these probability distributions (including their covariances with neighbouring grid boxes in space and time) allows one to define objectively stochastic-dynamic parametrisations, and hence representations of model uncertainty.

As an example of the potential for stochastic-dynamic parametrisation to be more complete in its representation of model uncertainty, consider the problem of parametrising deep convection. In bulk-formula representations of deep convection, it is assumed that the kinetic energy generated by the convective plumes (which in practice redistribute heat, moisture and momentum in response to a convectively unstable profile), is dissipated within the gridbox once the distribution has taken place. That is to say, whilst the magnitude of this dissipation will vary from parametrisation to parametrisation, all bulk-formula parametrisations in an MME will dissipate energy in the same way.

In practice (Lilly, 1983) in cases of organised deep convection, some of this initially divergent kinetic energy may project onto rotational modes and cascade upscale to resolved scales of climate models. Stochastic backscatter parametrisation takes a fraction of the energy that would otherwise be dissipated and injects it onto the resolved scale flow using a stochastic pattern generator. Stochastic backscatter is not specific to deep convection but also operates on diffusive and orographic gravity-wave drag parametrisations (Shutts, 2005). That is to say, the uncertainty associated with the extent to which small-scale kinetic energy can cascade upscale can be represented explicitly in stochastically parametrised ensembles, but is completely missing in an MME.

## 4.2 Systematic Error and Model Resolution

We quote from the IPCC Fourth Assessment Report (2007): "Models still show significant errors. Although these are generally greater at smaller scales, important large-scale problems also remain…The ultimate source of most such errors is that many important small-scale processes cannot be represented explicitly in models and so must be included in approximate form... This is partly due to limitations in computer power."

On the other hand increasing model resolution to try to reduce these biases is computationally demanding; increasing model resolution by a factor of 2 in each spatial direction implies an increase in computing cost by a factor of up to 16, allowing for the need for shorter time steps to maintain numerical stability. As a result, there is a growing belief in the need for dedicated computing infrastructure to be funded at the international level in order to support significant increases in model resolution (Shukla et al., 2008).

As model truncation scales decrease below 100 km, the slope of the energy spectrum of the atmosphere shallows from -3 to -5/3. This shallowing indicates that the realism of climate simulations may converge rather slowly to "truth" as resolution increases. For an isotropic homogeneous turbulent fluid, scaling and truncated model integrations suggest that there may actually be no convergence at all (Lorenz, 1969), and one of the Clay Mathematics Millennium Prize Problems (http://www.claymath.org/millennium/) is to prove (or disprove) this rigorously. In practice, since the accuracy of the effects of the representation of topography, the land/sea mask and other known forcings are manifestly improved with increased resolution, one would expect slow convergence in practice.

The very fact that convergence might be slow, suggests that stochastic-dynamic parametrisation might be considered a poor man's surrogate to high resolution particularly in Earth-System Models, where, for a given finite computational resource, the need to incorporate a full range of biogeochemical processes integrated over century and longer timescales is perhaps of higher priority than the desire (no matter how well justified) for high resolution.

The key reason why stochastic dynamic parametrisation may be a possible alternative to higher resolution is that, unlike the bulk-formula parametrisation, stochastic-dynamic parametrisation provides possible realisations of the sub-grid flow, rather than some inherently averaged bulk-formula. For example, in a cellular automaton model for deep convection, the individual cellular automata correspond to specific realisations of deep convective cloud systems.

Results from Jung et al. (2005) and Berner et al. (2008) have demonstrated the Cellular Automaton Stochastic Backscatter Scheme can indeed reduce systematic error, both in the tropics and in the extratropics.

## 4.3 Probabilistic CMOS for Stochastic Climate Models?

It is somewhat unfortunate that climate models, given their use in warning society of the dangers of profligate use of energy, need such energy-intensive hardware themselves. In this respect, there is another, albeit speculative, reason why future-generation climate models should be formulated stochastically.

In order to reach computing speeds in the petaflop range and above, high-performance computers are having to incorporate more and more processing elements, and in doing so are becoming more and more parallel. This situation has arisen because power density has become the dominant constraint in processor design, and as a result, the rate of increase in processor clock rates has not been sustainable. In order to compensate for this, the number of cores per chip has increased, and is estimated to double every 18 months in coming years.

Since stochastically-formulated models require Monte-Carlo type methods for their integration, such parallelisation does not itself pose an immediate problem of efficiency for a stochastic climate model. However, over and above this, there is a possible way in which stochastic climate models may be able to take advantage of new developments in computing technology which are being considered as a possible way to overcome this trend towards increased parallelism.

The transistor components in a digital computer's chip register a binary digit as electrons flow through the transistor in response to an applied voltage. Necessarily the movement of such electrons is "noisy" - indeed such noise is ultimately quantum mechanical and therefore inherently uncertain. In conventional computing, then to overcome this noise and ensure that the transistors register the correct bits, computer chips must run at relatively high voltage. As mentioned, the power requirements needed to maintain these high voltages is rapidly becoming the principal bottleneck to increasing computing speed and is the primary reason for the development of higher and higher levels of parallelism.

In current stochastic models, the random number generators are actually deterministic chaotic dynamical systems. This, and the power bottleneck mentioned, suggests an alternative approach: generate stochasticity from the electron flow itself.

Increasing the noise component in transistors can in principle be achieved by reducing the power to the chips, exactly what is required to increase basic CPU clock speed and maintain processing power. These ideas, whilst speculative, are not entirely fanciful. A prototype "probabilistic chip" has already been built (Kormaz et al, 2006; Palem, 2008) - and dubbed the PCMOS (Probabilistic Complementary Metal-Oxide Semiconductor) chip. The operating voltage of the logic circuits that calculate the least significant bits in a floating-point number is lowered. This procedure introduces stochasticity into individual calculations, but the underlying probability distributions are well defined. It is interesting to note that PCMOS technology has been applied to the problem of probabilistic cellular automata (Fuks, 2002), and could therefore be applied to the type of stochastic backscatter parametrisation scheme whose results are discussed in Section 4.

How precisely this type of technology could be utilised in a stochastic climate model clearly requires considerable further analysis. For example, could one envisage reducing operating voltages most for chips which perform calculations within the spatial domains of an ESM associated with the highest diagnosed numerical or parametrised dissipation?

In this respect, the idea that high-performance computer hardware should be customised for specific climate-prediction applications is not new (Wehner et al, 2008).

## 5. Conclusions

The primary conclusion of this paper is that next generation of climate models should be based on computational representations of the underlying equations of motion that are inherently stochastic.

- It is inherent to the scientific method that predictions must be accompanied by estimates of uncertainty. The multi-model method provides a pragmatic approach to the representation of uncertainty, but is *ad hoc*; the pool of world climate models has not been designed to span key uncertainties in the representation of unresolved processes, nor is it likely that one could design a multi-model system to do this in principle. Consistent with this, it has been shown that probability distributions of seasonal climate variables from multi-model ensembles are not accurate. Using seamless prediction techniques it has been shown that stochastic-dynamic methods are now beginning to challenge the multi-model ensemble as a technique for representing uncertainty in climate prediction.

- By the process of noise-induced rectification, which can occur in a nonlinear stochastic system even when the underlying probability distributions from which the stochastic processes are drawn from have zero mean, stochastic-dynamic parametrisations can (and has been shown to) reduce model systematic error. In this sense, stochastic representations of unresolved processes provide a poor man's surrogate to what would otherwise be a computationally demanding increase in model resolution. This is particularly relevant for Earth-System models which have to represent numerous physical and biogeochemical processes and have to be integrated over century or longer timescales. This is not to diminish the importance of the community's need for computational resources to improve the accuracy of climate model simulations, but rather stresses the pragmatic fact of the matter that resolution is not the only driver for such resources.

- Ultra high-performance computing is becoming increasing energy intensive and any potential capability to make use of computing technology with significantly lower energy demands should be exploited. Stochastic parametrisation may be able to take advantage of new low-energy computing technology provided by Probabilistic Complementary Metal-Oxide Semiconductor chips.

In short we propose here the notion of the Probabilistic Earth-System Model and look forward to its development and widespread use in future generation prediction systems.

## Acknowledgement

We acknowledge the ENSEMBLES project, funded by the European Commission's 6th Framework Programme through contract GOCE-CT-2003-505539.

# Figure captions:

Figure 1: Schematic showing the PDFs for a generic variable estimated from observations (a) and an MME (b). The dashed lines indicate observed tercile thresholds and the gray shaded area gives the frequency of exceeding the threshold (1/3 by definition). The same thresholds are used to estimate the corresponding MME frequency (shaded area in b). If the MME PDF were to be reliable, the area of the corresponding gray region should be 1/3. This is clearly not the case for the example shown; here the MME strongly overestimates the probability of the upper tercile.

Figure 2: a) Frequency of the IPCC AR4 multi-model ensemble for near-surface temperature in June - August (JJA) being in the lower observed tercile. White grid points correspond to points with frequencies between 24.7% and 42.5%, i.e. they are statistically not different to the reference frequency of 33.3% on the 99% significance level. Only 16% of the globe falls into this category. b): as a), but for December - February (DJF) being in the lower observed tercile. 19% of the globe falls into the 24.7%-42.5% range.

Figure 3: a) The attractor of the Lorenz (1963) SYSTEM in the X-Y plane. b) The attractor of the MODEL, obtained by rotating the SYSTEM attractor about an axis in the Z direction at the right regime centroids. c) Showing the response of both SYSTEM and MODEL to some prescribed forcing. The figure shows that linear bias correction cannot remedy the incorrect response of the MODEL to the imposed forcing, either in direction or magnitude.

Figure 4: Scatter plot of the Brier skill score of a) the stochastically parametrised ensemble, b) the calibrated stochastic-physics ensemble versus the nine-member multi-model ensemble and c) the combined stochastically parametrised/perturbed-parameters ensemble versus the 18-member multi-model ensemble. Each dot shows the skill score for the seasonal prediction of a specific parameter (two-metre temperature, precipitation, mean sea level pressure, 500-hPa geopotential height and 850-hPa temperature), start date (May in circles, November in triangles), lead time (one and three months), for three events (values above the upper tercile and the median, and below the lower tercile) over a region (tropical band, northern extratropics and southern extratropics) for a given pair of forecast systems.

Figure 1

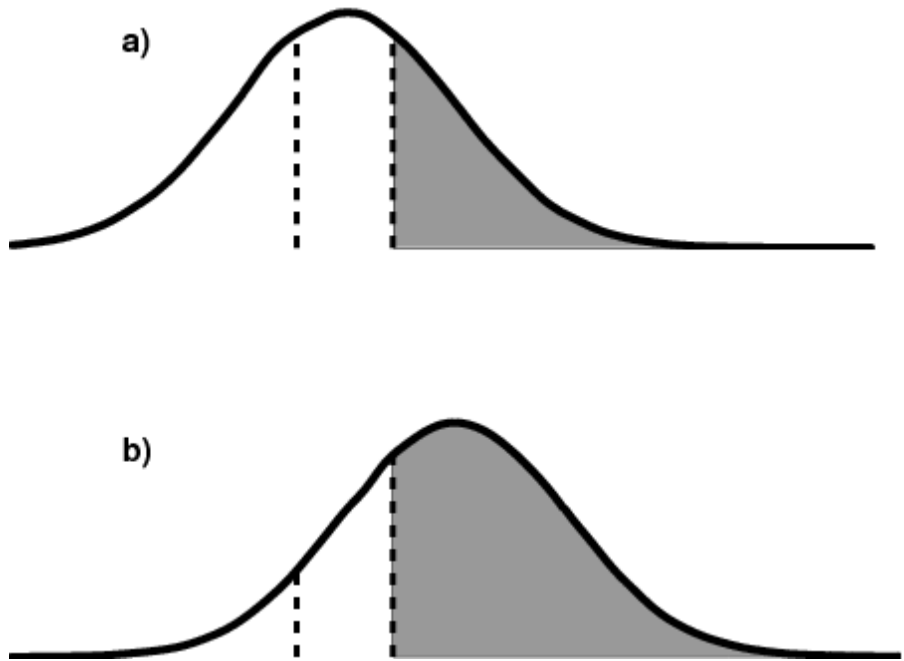

Schematic showing the PDFs for a generic variable estimated from observations (a) and an MME (b). The dashed lines indicate observed tercile thresholds and the gray shaded area gives the frequency of exceeding the threshold (1/3 by definition). The same thresholds are used to estimate the corresponding MME frequency (shaded area in b). If the MME PDF were to be reliable, the area of the corresponding gray region should be 1/3. This is clearly not the case for the example shown; here the MME strongly overestimates the probability of the upper tercile.

Figure 2

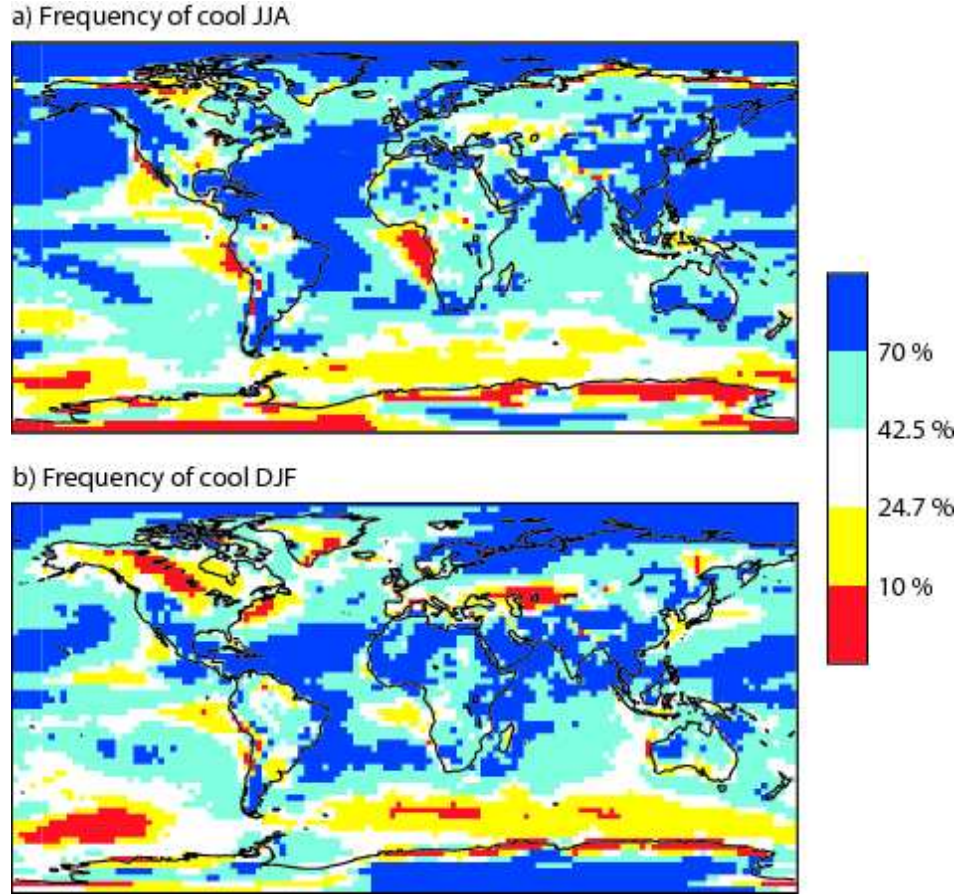

a) Frequency of the IPCC AR4 multi-model ensemble for near-surface temperature in June - August (JJA) being in the lower observed tercile. White grid points correspond to points with frequencies between 24.7% and 42.5%, i.e. they are statistically not different to the reference frequency of 33.3% on the 99% significance level. Only 16% of the globe falls into this category. b): as a), but for December - February (DJF) being in the lower observed tercile. 19% of the globe falls into the 24.7%-42.5% range.

Figure 3

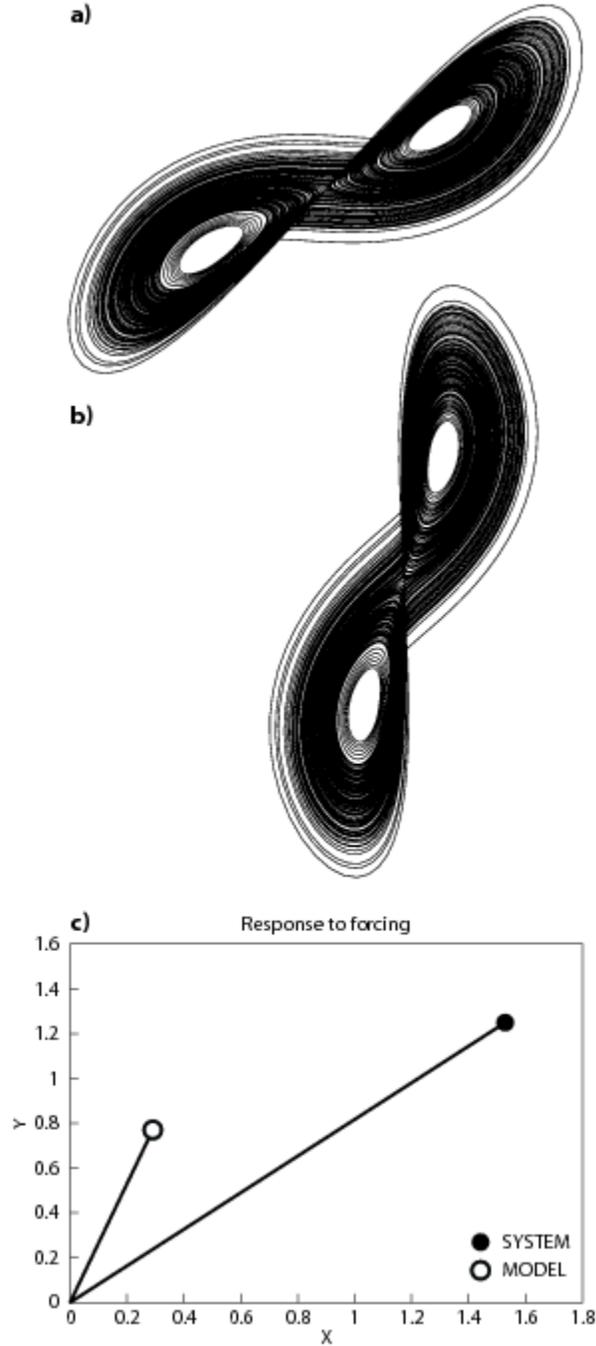

a) The attractor of the Lorenz (1963) SYSTEM in the X-Y plane. b) The attractor of the MODEL, obtained by rotating the SYSTEM attractor about an axis in the Z direction at the right regime centroids. c) Showing the response of both SYSTEM and MODEL to some prescribed forcing. The figure shows that linear bias correction cannot remedy the incorrect response of the MODEL to the imposed forcing, either in direction or magnitude.

Figure 4

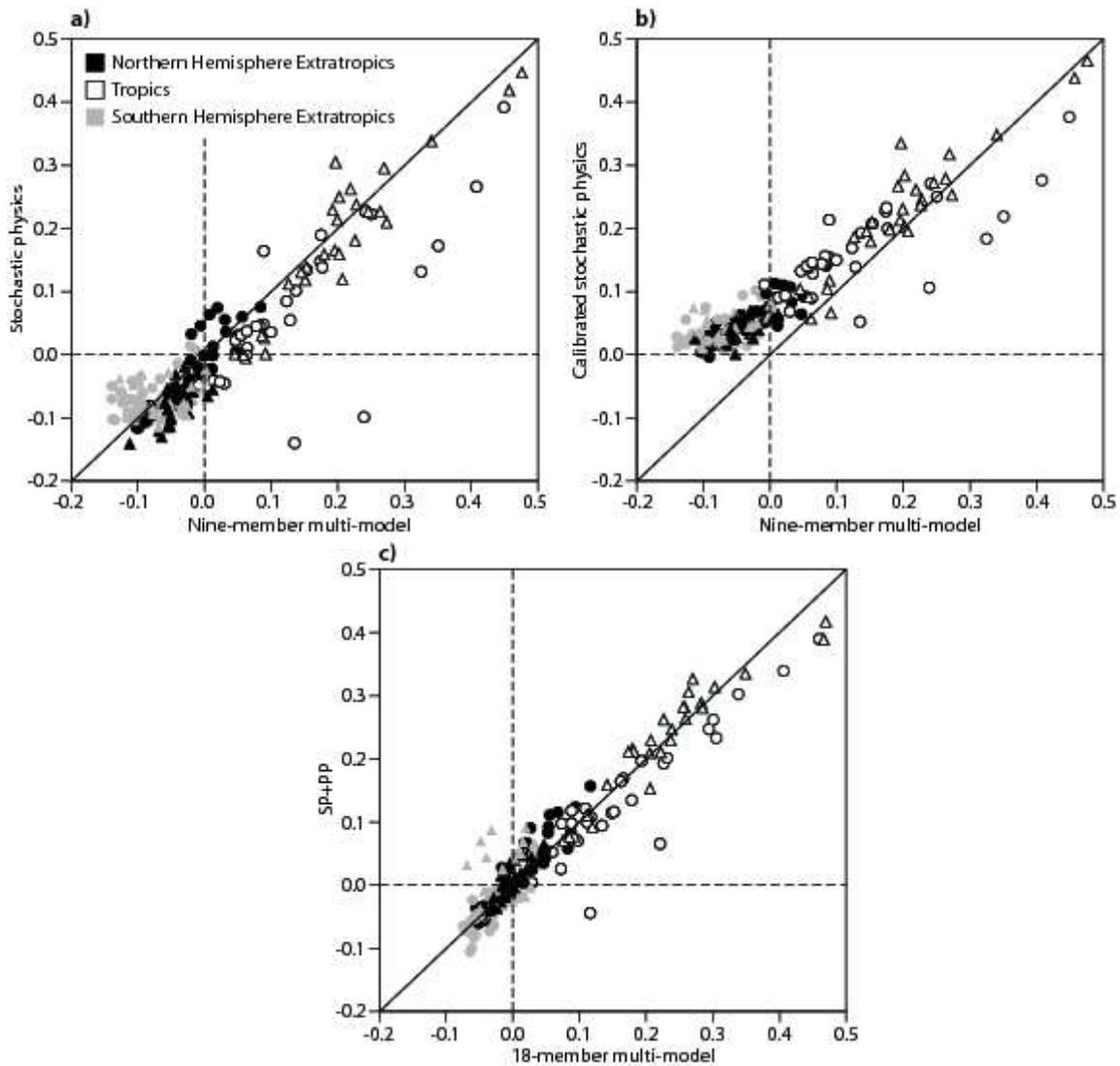

Scatter plot of the Brier skill score of a) the stochastically parametrised ensemble, b) the calibrated stochastic-physics ensemble versus the nine-member multi-model ensemble and c) the combined stochastically parametrised/perturbed-parameters ensemble versus the 18-member multi-model ensemble. Each dot shows the skill score for the seasonal prediction of a specific parameter (two-metre temperature, precipitation, mean sea level pressure, 500-hPa geopotential height and 850-hPa temperature), start date (May in circles, November in triangles), lead time (one and three months), for three events (values above the upper tercile and the median, and below the lower tercile) over a region (tropical band, northern extratropics and southern extratropics) for a given pair of forecast systems.